\documentclass[%
 reprint,
superscriptaddress,
 amsmath,amssymb,
 aps,
pra,
]{revtex4-2}

\usepackage{dcolumn}

\usepackage[colorlinks,citecolor=blue,linkcolor=red]{hyperref}
\usepackage{amsmath,amssymb,graphicx}

\begin{document}

\preprint{APS/123-QED}

\title{Bubbles and W-shaped solitons in Kerr media with fractional diffraction}

\author{Liangwei Zeng}
\affiliation{College of Physics and Optoelectronic Engineering, Shenzhen University, Shenzhen 518060, China}
\affiliation{Shenzhen Key Laboratory of Micro-Nano Photonic Information Technology,
Key Laboratory of Optoelectronic Devices and Systems of Ministry of Education and Guangdong Province,
College of Physics and Optoelectronic Engineering, Shenzhen University, Shenzhen 518060, China}

\author{Boris A. Malomed}
\email{\underline{malomed@tauex.tau.ac.il}}
\affiliation{Department of Physical Electronics, School of Electrical Engineering, Faculty of Engineering, and the Center for
Light-Matter Interaction, Tel Aviv University, P.O.B. 39040, Ramat Aviv, Tel Aviv, Israel}
\affiliation{Instituto de Alta Investigaci\'{o}n, Universidad de Tarapac\'{a}, Casilla 7D, Arica, Chile}

\author{Dumitru Mihalache}
\affiliation{Horia Hulubei National Institute of Physics and Nuclear Engineering, Magurele, Bucharest, RO-077125, Romania}

\author{Yi Cai}
\affiliation{College of Physics and Optoelectronic Engineering, Shenzhen University, Shenzhen 518060, China}
\affiliation{Shenzhen Key Laboratory of Micro-Nano Photonic Information Technology,
Key Laboratory of Optoelectronic Devices and Systems of Ministry of Education and Guangdong Province,
College of Physics and Optoelectronic Engineering, Shenzhen University, Shenzhen 518060, China}

\author{Xiaowei Lu}
\affiliation{College of Physics and Optoelectronic Engineering, Shenzhen University, Shenzhen 518060, China}
\affiliation{Shenzhen Key Laboratory of Micro-Nano Photonic Information Technology,
Key Laboratory of Optoelectronic Devices and Systems of Ministry of Education and Guangdong Province,
College of Physics and Optoelectronic Engineering, Shenzhen University, Shenzhen 518060, China}

\author{Qifan Zhu}
\affiliation{College of Physics and Optoelectronic Engineering, Shenzhen University, Shenzhen 518060, China}
\affiliation{Shenzhen Key Laboratory of Micro-Nano Photonic Information Technology,
Key Laboratory of Optoelectronic Devices and Systems of Ministry of Education and Guangdong Province,
College of Physics and Optoelectronic Engineering, Shenzhen University, Shenzhen 518060, China}

\author{Jingzhen Li}
\email{\underline{lijz@szu.edu.cn}}
\affiliation{College of Physics and Optoelectronic Engineering, Shenzhen University, Shenzhen 518060, China}
\affiliation{Shenzhen Key Laboratory of Micro-Nano Photonic Information Technology,
Key Laboratory of Optoelectronic Devices and Systems of Ministry of Education and Guangdong Province,
College of Physics and Optoelectronic Engineering, Shenzhen University, Shenzhen 518060, China}

\date{April 9, 2021}

\begin{abstract}
We demonstrate that, with the help of a Gaussian potential barrier, dark modes in the form of a local depression (``bubbles") can be supported by the repulsive Kerr nonlinearity in combination with fractional dimension. Similarly, W-shaped modes are supported by a double potential barrier. Families of the modes are constructed in a numerical form, and also by means
of the Thomas-Fermi and variational approximations. All these modes are stable, which is predicted by computation of eigenvalues for small perturbations and confirmed by direct numerical simulations.
\end{abstract}

\maketitle

\section{Introduction}

Bright, dark, and gray solitons are fundamental modes that are known in various forms in nonlinear dispersive media \cite{book1,book2,book3,LPP5,REV1,Rev4,REV7}, such as optical waveguides \cite{Abd-book,Hasegawa,DS1,DS2,REV2,Maimistov,REV6,REV3,MalomedMihalache2019} and Bose-Einstein condensates \cite{Hulet,Abdullaev,Bagnato,Salasnich,Harko,Passos-ND}. Dark solitons, with zero-crossing shapes, are quiescent modes \cite{DS1,DS2,Frantz}, while in the moving state they assume a complex (gray) shape, with a nonzero minimum of the absolute value \cite{GRAY1,GRAY2,GRAY3,GRAY4}. Dark and gray solitons carry a topological charge, represented by the opposite signs of the finite-amplitude background fields on the two sides of these solitons. Nonlinear Schr\"{o}dinger equations (NLSEs), with the attractive or repulsive sign of the cubic nonlinearity, are universal models producing soliton families of bright and dark types.

Varieties of fundamental and excited soliton states may be essentially expanded in spatially inhomogeneous settings \cite{REV7,INH1,INH3,INH4,INH5,INH6}. One of methods commonly used to create inhomogeneous nonlinear media is the introduction of linear lattice potentials \cite{LPP5,LPP1,3Din2D,LPP2,LPP3,LPP4,LPP6}, which support families of gap solitons, such as fundamental \cite{LL1}, dipole \cite{LL2} and vortex ones \cite{Kevrekid,LL3,INH2,LL4,LL5}. Another setup is based on the use of nonlinear lattices, i.e., spatially periodic modulations of the nonlinearity coefficient, which also help to build diverse soliton families \cite{NLREV}, including fundamental, multipole, and vortical ones \cite{NL1,NL2,NL3}. Furthermore, the self-repulsive cubic term, with the local strength growing fast enough from the center to periphery, makes it possible to maintain exceptionally robust species of self-trapped states \cite{SDN1,SDN2,SDN3}, such as solitary vortices \cite{SDN4,SDN5,SDN6,SDN7}, soliton gyroscopes \cite{SDN10}, \textit{hopfions} (twisted vortex rings) \cite{SDN11}, vortex clusters \cite{SDN12,SDN13}, and flat-top solitons \cite{SDN8,SDN9}.

Depending on the form of the self-repulsive nonlinearity, dark modes also may exist in the form of \textquotedblleft bubbles", i.e., quiescent real states with a nonzero local minimum of the amplitude \cite{bubble1,bubble2,bubble3}. Unlike dark solitons, bubbles do not carry a topological charge.

A particular species of spatial modes produced by NLSEs is represented by W-shaped solitons, so named because of their shape with two local minima \cite{WSS1,WSS2,WSS3,WSS4,WSS5,WSS6,WSS7}. Physical settings that support W-shaped solitons include nonlinear fibers with higher-order effects \cite{WSS1,WSS2}, optical materials with self-induced transparency \cite{WSS3}, negative-refractive-index materials \cite{WSS6}, as well as the media with inhomogeneous repulsive nonlinearity \cite{WSS7}.

An essential extension of the class of Schr\"{o}dinger equations was introduced by Laskin \cite{Lask1,Lask2,Lask3,Lask4}, in the form of the fractional Schr\"{o}dinger equation, which includes the fractional derivative, instead of the usual second-order differential term. The model was developed as one for a quantum particle moving by way of \textit{L\'{e}vy flights} (random jumps). The addition of the cubic term to the equation leads to the concept of the nonlinear fractional Schr\"{o}dinger \, equation (NLFSE). In the context of the quantum many-body theory, NLFSE\ may be considered as an appropriate equation of the Gross-Pitaevskii (mean-field) type for a Bose-Einstein condensate of such particles \cite{Belic',Frac-BEC1}. Further, NLFSE has drawn much interest due to possibilities of its experimental realization in the classical form, in condensed matter physics \cite{EXP1,EXP2} and optics \cite{EXP3}. Solitons produced by NLFSEs have been predicted in various forms \cite{Frac1,Frac2,Frac4,Frac5,Frac6,Frac7,Frac8,Frac9,Frac10,Frac12,Frac13,Frac14,Frac15,Frac16,Frac17,Frac18,Molina,Frac19}, including ``accessible" so\-li\-tons \cite{Frac1,Frac2} and ones supported by nonlinear lattices \cite{Frac4,Frac19}, symmetry-breaking states \cite{Frac6,Frac7,Frac8,Frac9}, gap solitons \cite{Frac10}, solitary vortices \cite{Frac12,Frac13}, multi-peak modes \cite{Frac14}, soliton clusters \cite{Frac15,Frac16}, coupled solitons \cite{Frac17,Frac18}, discrete ones
\cite{Molina}, and localized pulses in fractional complex Ginzburg-Landau equation \cite{CGL}.

Although many soliton species have been found as solutions of NLFSEs, bubble-type and W-shaped modes were not yet addressed in the case of equations of the fractional order. Such states, supported by the repulsive cubic nonlinearity, are the subject of the present work. We demonstrate that they can be stabilized by a Gaussian potential barrier (or two barriers, for the W-shaped states) added to the setting.

The rest of the paper is organized as follows. In Sec. \ref{sec2}, we present the model and formulate the linear stability analysis for it. The same section includes two analytical methods, \textit{viz}., a crude Thomas-Fermi (TF) approximation and a more accurate variational approach. Numerical results for the bubbles and W-shaped modes are reported in two
parts of Sec. \ref{sec3}. The paper is concluded by Sec. \ref{sec4}.

\section{The model and analytical approximations}

\label{sec2}

\subsection{Basic equations}

The NLFSE is introduced in terms of the spatial-domain light propagation, with field amplitude $E$, along axis $z$ in a waveguide with the defocusing Kerr nonlinearity, under the action of fractional diffraction in the transverse direction, $x$. In the scaled form, the equation takes the form of
\begin{equation}
i\frac{\partial E}{\partial z}=\frac{1}{2}\left( -\frac{\partial ^{2}}{\partial x^{2}}\right) ^{\alpha /2}E+V(x)E+4\left\vert E\right\vert ^{2}E,
\label{NLFSE}
\end{equation}
where $(-\partial ^{2}/\partial x^{2})^{\alpha /2}$ is the fractional Laplacian, determined by the L\'{e}vy index $\alpha $. In fact, the fractional derivative is defined as an integral operator,
\begin{equation}
\left( -\frac{\partial ^{2}}{\partial x^{2}}\right) ^{\alpha /2}E=\frac{1}{2\pi }\int_{-\infty }^{+\infty }dp|p|^{\alpha }\int_{-\infty }^{+\infty}d\xi e^{ip(x-\xi )}E(\xi ).
\label{int}
\end{equation}
The strength of the self-repulsive nonlinearity in Eq. (\ref{NLFSE}) is fixed to be $4$ by means of an obvious rescaling of $E$.

The implementation of the fractional diffraction, modeled by Eq. (\ref{int}), in optics was elaborated by Longhi \cite{EXP3}. It can be realized in a Fabry-Perot resonator, into which two lenses and two phase masks are inserted, with the phase shift $f(x)$, introduced by the mask, which depends on the transverse coordinate, $x$, as $f(x)\sim |x|^{\alpha }$. In that case, the variable $z$ in Eq. (\ref{NLFSE}) is proportional to the number of circulations of light beams in the resonator. It is also relevant to mention early works \cite{early1} and \cite{early2}, which proposed the implementation of the fractal diffraction by means of optical filters.

In works on the NLFSE with the attractive cubic term, which corresponds to the negative coefficient in front of the cubic term in Eq. (\ref{NLFSE}), the L\'{e}vy index takes the values $1<\alpha \leq 2$, because at $\alpha \leq 1$ such an equation gives rise to collapse, hence all stationary solutions are unstable. On the other hand, solitons may be stabilized by a
self-repulsive quintic term added to the NLFSE \cite{Frac12}, as well as by a trapping potential \cite{Frac14}, which makes it possible to obtain stable solutions at $\alpha \leq 1$ (e.g., for $\alpha =1$ \cite{Frac12} and $0.7<\alpha \leq 1$ \cite{Frac14}).

The potential in Eq. (\ref{NLFSE}), which supports bubble modes, is adopted in the form of a Gaussian barrier, with height $A$ and width $W$:
\begin{equation}
V(x)=A\exp \left( -\frac{x^{2}}{2W^{2}}\right) .  \label{GSL}
\end{equation}
In optics, it can be readily realized by means of an inhomogeneity of the refractive index. Note that, while potential barriers repel bright solitons, they attract dark modes. Accordingly, to produce W-shaped states, the potential will be taken as a pair of Gaussian barriers,
\begin{equation}
V(x)=A\left\{ \exp \left[ -\frac{(x-x_{0})^{2}}{2W^{2}}\right] +\exp \left[ -\frac{(x+x_{0})^{2}}{2W^{2}}\right] \right\} ,
\label{WSL}
\end{equation}
where the constant $x_{0}>0$ controls the separation between them.

Stationary states with real propagation constant $k$ are looked for as solution to Eq. (\ref{NLFSE}) in the form of
\begin{equation}
E(x,z)=U(x)\mathrm{exp}(ikz).
\label{k}
\end{equation}
The substitution of this expression in Eq. (\ref{NLFSE}) leads to an equation for real function $U(x)$:
\begin{equation}
-kU=\frac{1}{2}\left( -\frac{\partial ^{2}}{\partial x^{2}}\right) ^{\alpha/2}U+V(x)U+4U^{3}.
\label{NLFSES}
\end{equation}
In particular, dark modes are supported by the modulationally stable flat background, represented by constant solutions of Eq. (\ref{NLFSES}) with $V(x)=0$,
\begin{equation}
U_{\mathrm{b}}=\sqrt{-k}/2,  \label{Ub}
\end{equation}
which exist for all negative values of $k$.

Note that Eq. (\ref{NLFSES}), combined with the definition of fractional-diffraction operator (\ref{int}), is invariant with respect to rescaling,
\begin{gather}
\tilde{x}\equiv X_{0}^{-1}x,\tilde{k}\equiv X_{0}^{\alpha }k,  \notag \\
\tilde{V}(\tilde{x})\equiv X_{0}^{\alpha }V(X_{0}\tilde{x}),\tilde{U}(\tilde{x})\equiv X_{0}^{\alpha /2}U\left( X_{0}\tilde{x}\right),
\label{x0}
\end{gather}
with arbitrary scaling factor $X_{0}$. The invariance makes it possible to fix either amplitude or width of the potential barrier (\ref{GSL}) (similarly, one can fix one parameter for the double barrier (\ref{WSL})). Nevertheless, presenting numerical results below, we independently vary the constants $k$, $A$, and $W$, as the variation of any of them is a natural scenario for experiments.

Equation (\ref{NLFSE}) conserves the total norm, which represents the integral power of the light beam in the underlying optical model:
\begin{equation}
P=\int_{-\infty }^{+\infty }\left\vert E(x)\right\vert ^{2}dx.
\label{P}
\end{equation}
For stationary dark modes, $P$ diverges because of the contribution of the nonvanishing background, while an appropriate characteristic of the modes is the \textit{power defect},
\begin{equation}
\Delta P\equiv \int_{-\infty }^{+\infty }\left[ U_{\mathrm{b}}^{2}-U^{2}(x)\right] dx,
\label{POWER}
\end{equation}
where $U_{\mathrm{b}}^{2}$ is the background density supporting the dark states, given by Eq. (\ref{Ub}).

\subsection{The Thomas-Fermi (TF) approximation}

The simplest analytical approximation for stationary modes, which, actually, completely ignores the fractional dimension, dropping the (fractional) derivative term in Eq. (\ref{NLFSES}), is the TF approximation, which is widely applied to the consideration of Bose-Einstein condensates with the self-repulsive nonlinearity \cite{Pit}. In this approximation, Eq. (\ref{NLFSES}) yields
\begin{equation}
U_{\mathrm{TF}}^{2}(x)=(1/4)\left[ |k|-V(x)\right],
\label{TFA}
\end{equation}
provided that the maximum value of the potential is subject to condition $V_{\max }<|k|$ (recall that we consider $k<0$). In the case of $V_{\max}>|k| $, the TF approximation takes the form including an empty (zero-field) segment:
\begin{equation}
U_{\mathrm{TF}}^{2}(x)=\left\{
\begin{array}{c}
(1/4)\left[ |k|-V(x)\right] ,~\mathrm{at}~V(x)<|k|, \\
0,\mathrm{at}~V(x)>|k|.
\end{array}
\right.
\label{TFA2}
\end{equation}
In the case of potential barrier (\ref{GSL}), the empty ``hole" is the central segment, $|x|<W\sqrt{2\ln \left( A/|k|\right) }$. However, numerically found profiles of the bubbles, displayed below (see Figs. \ref{fig2}(a,d,g)), do not feature any \textquotedblleft hole", even in an approximate form. Thus, the fractional diffraction operator supports profiles that are quite different from their TF counterparts.

The TF approximation produces the corresponding values of the power defect, $\Delta P$. In particular, in the case of $A<|k|$, the substitution of expression (\ref{TFA}), with $V(x)$ taken as per Eq. (\ref{GSL}), yields a simple expression, which does not depend on $k$:
\begin{equation}
\left( \Delta P_{\mathrm{TF}}\right) _{A<|k|}=\sqrt{\pi /8}AW.
\label{DeltaP-TF}
\end{equation}
In the case of the double potential barrier (\ref{WSL}), the TF approximation yields a result that is twice as large, \textit{viz}.,
\begin{equation}
\left( \Delta P_{\mathrm{TF}}\right) _{\mathrm{double}}=\sqrt{\pi /2}AW.
\label{double}
\end{equation}
It is shown below in Figs. \ref{fig2}(c,f) that, although the TF prediction of the bubbles' shapes is not accurate, the expression (\ref{DeltaP-TF}) correctly predicts essential features of the dependence of $\Delta P$ on parameters.

\begin{figure}[tbp]
\begin{center}
\includegraphics[width=1.0\columnwidth]{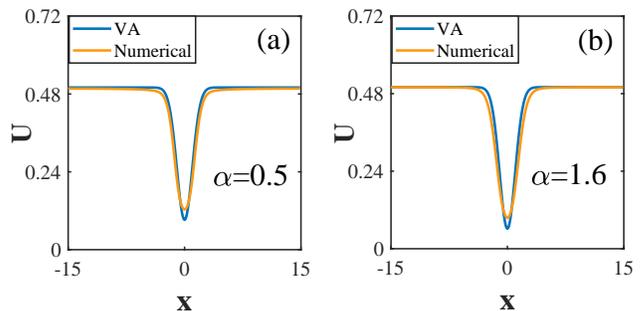}
\end{center}
\caption{Numerically obtained profiles of bubbles and their variational counterparts (yellow and blue lines, respectively), supported by the potential barrier (\protect\ref{GSL}), for values of the L\'{e}vy index $\protect\alpha =0.5$ (a) and $1.6$ (b). Parameters of the potential barrier are $A=2$, $W=1$, and $k=-1$.}
\label{fig1}
\end{figure}

\subsection{The variational approximation}

For a fixed value of $k$ and, accordingly, the background value of $U$ fixed as per Eq. (\ref{Ub}), the bubble solution may be looked for as
\begin{equation}
U(x)=\left( \sqrt{-k}/2\right) -u(x),
\label{-u}
\end{equation}
with a positive localized field $u(x)$ satisfying the complimentary equation,
\begin{gather}
\frac{1}{2}\left( -\frac{\partial ^{2}}{\partial x^{2}}\right) ^{\alpha/2}u+\left( V(x)-2k\right) u-6\sqrt{-k}u^{2}+4u^{3} \notag \\
=\frac{\sqrt{-k}}{2}V(x).
\label{u}
\end{gather}
The fact that the Eq. (\ref{u}) can be derived from the Lagrangian,
\begin{gather}
L=\frac{1}{8\pi }\int_{-\infty }^{+\infty }dp|p|^{\alpha }\int_{-\infty}^{+\infty }dx\int_{-\infty }^{+\infty }d\xi e^{ip(x-\xi )}u(\xi )u(x) \notag \\
+\int_{-\infty }^{+\infty }dx\left[ \left( \frac{V}{2}-k\right) u^{2}-2\sqrt{-k}u^{3}+u^{4}-\frac{\sqrt{-k}V}{2}u\right],
\label{L}
\end{gather}
where the definition (\ref{int}) for the fractional derivative is used, suggests one to apply the variational approximation, similar to how it was used for bright solitons of the NLFSE with the cubic self-attraction in Ref. \cite{CGL}. A natural variational ansatz for the bubble is a Gaussian with amplitude $a>0$ considered as a free parameter:
\begin{equation}
u_{\mathrm{ans}}(x)=a\exp \left( -\frac{x^{2}}{2W^{2}}\right).
\label{ans}
\end{equation}
A more general ansatz can be used too, with its width treated as another free parameter, rather than fixed to be equal to $W$, but the respective generalization leads to quite cumbersome variational equations.

\begin{figure*}[tbp]
\begin{center}
\includegraphics[width=1.35\columnwidth]{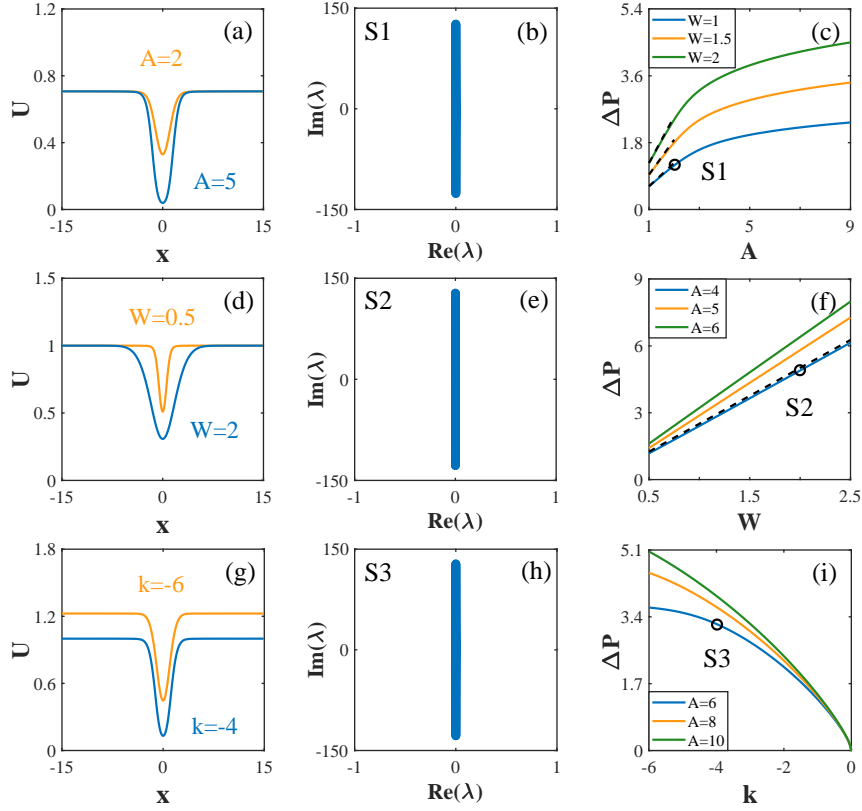}
\end{center}
\caption{(a) Profiles of typical bubble solutions supported by the potential barrier (\protect\ref{GSL}) with different values of $A$ and fixed parameters $W=1$, $k=-2$. (b) Eigenvalues of small perturbations around the bubble with $A=2$, $W=1$, $k=-2$. (c) Power defect $\Delta P$ of the bubble solutions vs. amplitude $A$ of the potential barrier, see Eqs. (\protect\ref{POWER}) and (\protect\ref{DeltaP-TF}), for different values of width $W$ at a fixed value of the propagation constant, $k=-2$. The black dashed lines in panels (c,f) are the TF predictions, as given by Eq. (\protect\ref{DeltaP-TF}). (d) Bubble profiles with different values of $W$ at $A=4$, $k=-4$. (e) Eigenvalues of small perturbations at $A=4$, $W=2$, $k=-4$. (f) $\Delta P$ vs. $W$ for different values of $A$ at $k=-4$. (g) Bubble profiles with different values of $k$ at $A=6$, $W=1$. (h) Eigenvalues of small perturbations at $A=6$, $W=1$, $k=-4$. (i) $\Delta P$ vs. $k$ for different values of $A$ at $W=1$. The evolution of perturbed propagations of the bubbles marked by S1--S3 is displayed in Figs. \protect\ref{fig4}(a--c), respectively. Except for the left panel of Fig. \protect\ref{fig1}, the L\'{e}vy index $\protect\alpha $ in Eq. (\protect\ref{NLFSE}) is fixed as $\protect\alpha =1.6$.}
\label{fig2}
\end{figure*}

The substitution of ansatz (\ref{ans}) in Lagrangian (\ref{L}) and subsequent integration yields the effective Lagrangian:
\begin{gather}
\frac{1}{\sqrt{\pi }W}L_{\mathrm{eff}}=-\frac{\sqrt{-k}}{2}Aa-\sqrt{-\frac{8k}{3}}a^{3}+\frac{1}{\sqrt{2}}a^{4} \notag \\
+\left[ \frac{\Gamma \left( \left( \alpha +1\right) /2\right) }{4\sqrt{\pi }}W^{-\alpha }-k+\frac{A}{\sqrt{6}}\right] a^{2},
\label{Leff}
\end{gather}
where $\Gamma $ is the Euler's Gamma-function. The value of the amplitude $a$ in ansatz (\ref{ans}) is predicted by the Euler-Lagrange equation, $dL_{\mathrm{eff}}/da=0$, which takes the form of
\begin{gather}
\sqrt{2}a^{3}-\sqrt{-6k}a^{2}+\left[ \frac{\Gamma \left( \left( \alpha+1\right) /2\right) }{4\sqrt{\pi }}W^{-\alpha }-k+\frac{A}{\sqrt{6}}\right] a \notag \\
-\frac{\sqrt{-k}}{4}A=0.
\label{dLda}
\end{gather}
With $a\allowbreak $ found as a solution of this cubic equation, one can produce the variational prediction for the power defect, substituting expressions (\ref{-u}) and (\ref{ans}) in Eq. (\ref{POWER}):
\begin{equation}
\Delta P_{\mathrm{var}}=2\sqrt{-\pi k}Wa-\sqrt{\pi }Wa^{2}.
\label{DeltaP-var}
\end{equation}

\begin{figure*}[tbp]
\begin{center}
\includegraphics[width=1.35\columnwidth]{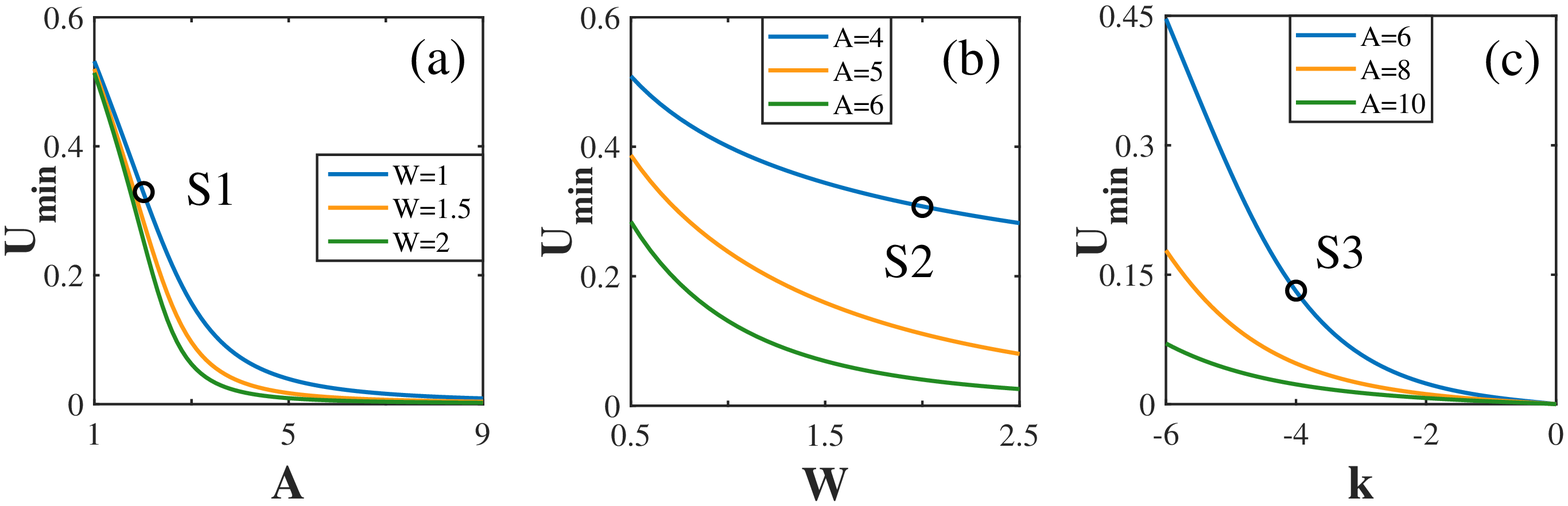}
\end{center}
\caption{(a) The minimal value $U_{\mathrm{min}}$ of the bubble solution, supported by the potential barrier (\protect\ref{GSL}), vs. $A$ for different values of $W$ at $k=-2$. (b) $U_{\mathrm{min}}$ vs. $W$ for different values of $A$ at $k=-4$. (c) $U_{\mathrm{min}}$ vs. $k$ for different values of $A$ at $W=1$.}
\label{fig3}
\end{figure*}

\begin{figure*}[tbp]
\begin{center}
\includegraphics[width=1.35\columnwidth]{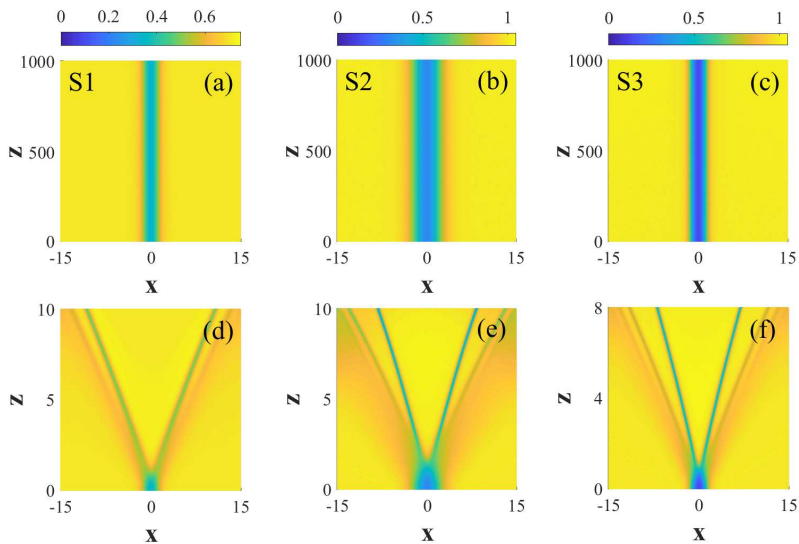}
\end{center}
\caption{Stable propagation of perturbed bubbles supported by the potential barrier (\protect\ref{GSL}): (a) at $A=2$, $W=1$, $k=-2$; (b) at $A=4$, $W=2$, $k=-4$; (c) at $A=6$, $W=1$, $k=-4$. Panels (d,e,f) display diffraction of the same inputs as in (a,b,c), respectively, but without the potential barrier ($A=0$).}
\label{fig4}
\end{figure*}

\subsection{The formulation of the stability analysis}

Stability of the stationary modes was investigated, as usually, by taking perturbed ones as
\begin{equation}
E=[U(x)+p(x)\mathrm{exp}(\lambda z)+q^{\ast }(x)\mathrm{exp}(\lambda ^{\ast}z)]\mathrm{exp}(ikz),  \label{PERB}
\end{equation}
where $p(x)$ and $q^{\ast }(x)$ are eigenmodes of small perturbations with instability growth rate $\lambda $, the asterisk standing for complex conjugate \cite{LPP5}. Substituting this ansatz in Eq. (\ref{NLFSE}) and linearization leads to the eigenvalue problem based on equations
\begin{equation}
\left\{
\begin{aligned}
i\lambda p=+\frac{1}{2}\left(-\frac{\partial^2}{\partial x^2}\right)^{\alpha/2}p+(k+V)p+4U^2(2p+q),\\
i\lambda q=-\frac{1}{2}\left(-\frac{\partial^2}{\partial x^2}\right)^{\alpha/2}q-(k+V)q-4U^2(2q+p).
\end{aligned}
\right.
\label{LAS}
\end{equation}
(in the theory of Bose-Einstein condensates, they are called Bogoliubov-de Gennes equations \cite{Pit}). The stationary solution is stable if all the eigenvalues $\lambda $ produced by Eq. (\ref{LAS}) are pure imaginary ones.

\section{Numerical results}

\label{sec3}

Numerical solutions of Eq. (\ref{NLFSES}) were constructed by means of the modified squared-operator method \cite{LPP5}. Stability and instability domains for families of stationary states were identified by means of eigenvalues produced by a numerical solution of Eq. (\ref{LAS}) and verified by direct numerical simulations of Eq. (\ref{NLFSE}) for the perturbed evolution of the same states, performed with the help of the split-step fast-Fourier-transform method. The numerical findings are presented below, chiefly, for $\alpha =1.6$, as a characteristic value of the L\'{e}vy index that is essentially different from $\alpha =2$, which corresponds to the usual NLSE. To illustrate the generality of the obtained results, some of them are additionally displayed, in Fig. \ref{fig1}, for a very different value of the L\'{e}vy index, $\alpha =0.5$.

\begin{figure*}[tbp]
\begin{center}
\includegraphics[width=1.65\columnwidth]{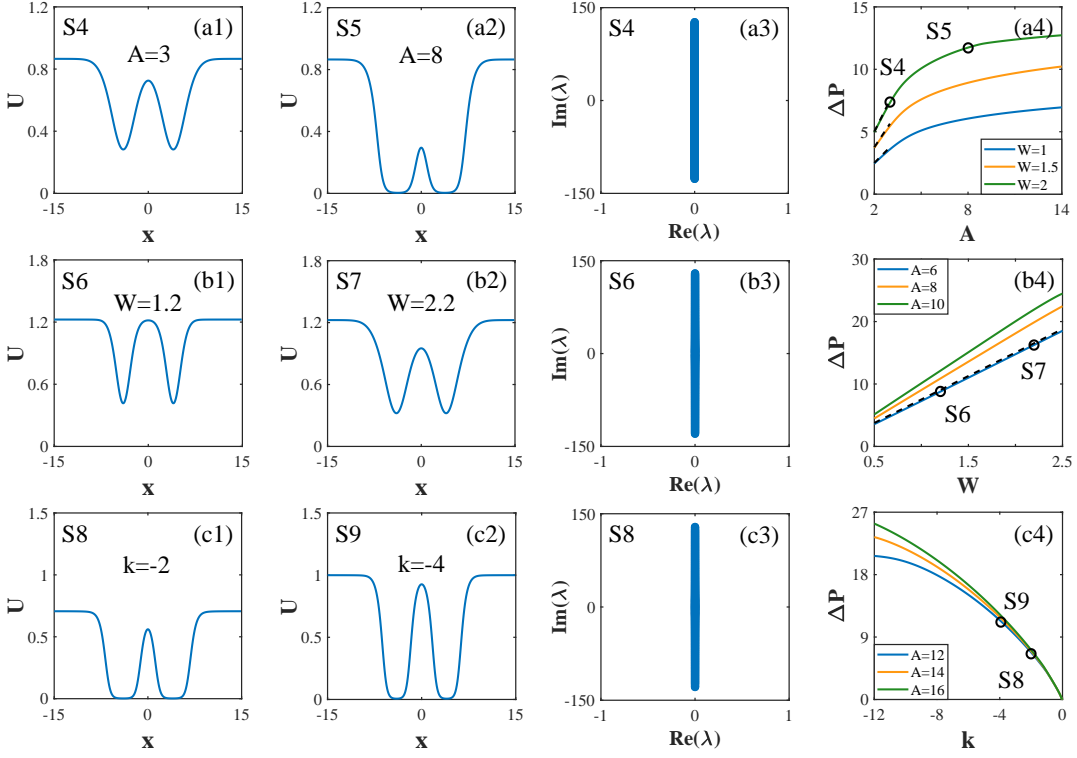}
\end{center}
\caption{The same as in Fig. \protect\ref{fig2}, but for W-shaped dark modes supported by the double potential barrier (\protect\ref{WSL}). Profiles of the modes at $W=2$, $k=-3$: (a1) with $A=3$; (a2) with $A=8$. (a3) The spectrum of perturbation eigenvalues for the W-shaped mode at $A=3$, $W=2$, $k=-3$. (a4) $\Delta P(A)$ with different values of $W$ at $k=-3$. The black dashed lines in panels (a4,b4) are the TF predictions (\protect\ref{double}). Profiles of the mode at $A=6$, $k=-6$: (b1) for $W=1.2$; (b2) for $W=2.2$. (b3) The spectrum of eigenvalues for $A=6$, $W=1.2$, $k=-6$. (b4) $\Delta P(W)$ for different values of $A$ at $k=-6$. Profiles of the modes at $A=12$, $W=1.4$: (c1) for $k=-2$; (c2) for $k=-4$. (c3) The spectrum of perturbation eigenvalues for $A=12$, $W=1.4$, $k=-2$. (c4) $\Delta P(k)$ for different values of $A$ at $W=1.4$. The propagations of perturbed modes marked by S4, S6, and S8 are displayed in Figs. \protect\ref{fig7}(a--c), respectively. In this and the other figures, $x_{0}=4$ is fixed in Eq. (\protect\ref{WSL}).}
\label{fig5}
\end{figure*}

\begin{figure*}[tbp]
\begin{center}
\includegraphics[width=1.35\columnwidth]{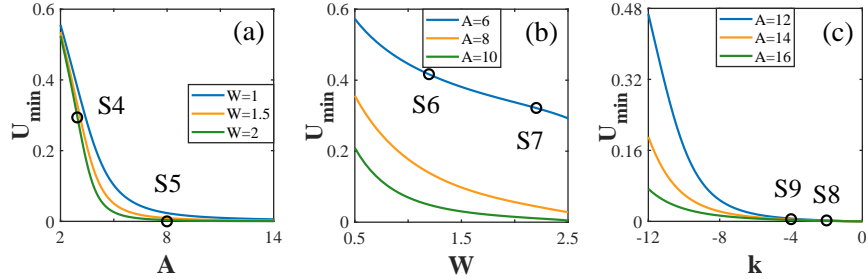}
\end{center}
\caption{The same as in Fig. \protect\ref{fig3}, but for the W-shaped modes. (a) $U_{\mathrm{min}}$ of the modes versus $A$ for different values of $W$ at $k=-3$. (b) $U_{\mathrm{min}}$ versus $W$ for different values of $A$ at $k=-6$. (c) $U_{\mathrm{min}}$ versus $k$ for different values of $A$ at $W=1.4$.}
\label{fig6}
\end{figure*}

\begin{figure*}[tbp]
\begin{center}
\includegraphics[width=1.35\columnwidth]{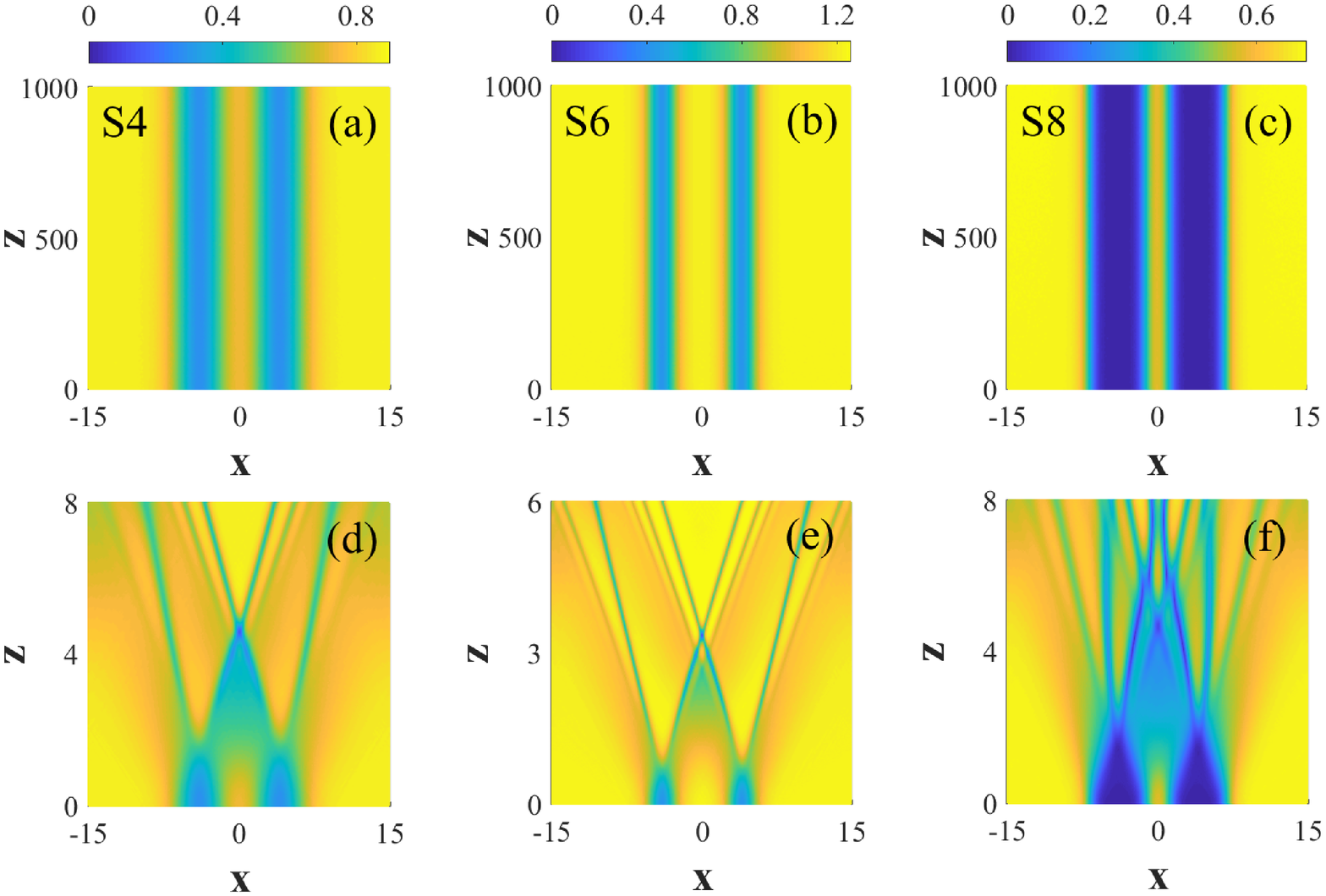}
\end{center}
\caption{The stable propagation of perturbed W-shaped dark modes: (a) at $A=3$, $W=2$, $k=-3$; (b) at $A=8$, $W=1.2$, $k=-6$; (c) at $A=12$, $W=1.4$, $k=-2$. (d,e,f) Diffraction of the same inputs in the case when the double Gaussian potential barrier is removed.}
\label{fig7}
\end{figure*}

\subsection{Bubbles}

\label{sec3a}

Typical profiles of bubbles for the above-mentioned values of the L\'{e}vy index, $\alpha =0.5$ and $1.6$, are displayed in Fig. \ref{fig1}, along with their counterparts predicted by the variational approximation, as per Eqs. (\ref{-u}), (\ref{ans}), and (\ref{dLda}). It is seen that the approximation predicts the profiles with good accuracy.

Typical profiles of bubbles, spectra of eigenvalues of small perturbations around them, and characteristics of the bubble-solution families, in the form of the dependence of the respective power defect (\ref{POWER}) on the parameters, are displayed in Fig. \ref{fig2}. In particular, the profile and eigenvalue spectrum shown in panels (a) and (b) represent the bubble marked
by S1 in (c). Note that $\Delta P$ increases with $A$ in panel (c), at different values of width $W$ of the Gaussian potential barrier (\ref{GSL}). This dependence agrees with panel (a), where the minimum level $U(x=0)\equiv U_{\mathrm{min}}$ of the bubble's field decreases, quite naturally, with the increase of $A$.

Further, the dependence $\Delta P(W)$ for different values of $A$ is displayed in Fig. \ref{fig2}(f), which shows that $\Delta P$ increases with $W$ almost linearly for different values of $A$. This property agrees with panel (d), which shows that, also naturally, $U_{\mathrm{min}}$ decreases with the increase of width $W$ of the repulsive barrier, and the bubble's width increases, following the increase of $W$. The profile and eigenvalue spectrum shown in panels (d) and (e) of Fig. \ref{fig2} represent the bubble marked by S2 in Fig. \ref{fig2}(f). Note that the nearly linear dependence $\Delta P(A)$ in the region of $A<|k|$ in Fig. \ref{fig2}(c), and the linear dependence $\Delta P(W)$ in Fig. \ref{fig2}(f) are consistent with the simple TF prediction, given by Eq. (\ref{DeltaP-TF}).

The dependence $\Delta P(k)$ is presented in Fig. \ref{fig2}(i), which demonstrates that $\Delta P$ increases with the increase of $|k|$ for different values of $A$. The respective typical profiles of the bubbles with different $k$\ are plotted in Fig. \ref{fig2}(g), showing that $U_{\min }$ increases with $|k|$, while the bubble's width remains nearly constant. The profile and eigenvalue spectrum shown in panels (g) and (h) of Fig. \ref{fig2} represent the bubble marked by S3 in Fig. \ref{fig2}(i).

The minimum value $U_{\mathrm{\min }}$ being an essential characteristic of the bubble, we display its dependence on $A$ for different values of $W$ in Fig. \ref{fig3}(a), from which one can clearly see that (as mentioned above) $U_{\mathrm{\min }}$ decreases with the increase of $A$. The same property is predicted by the TF approximation, as long as $A$ remains smaller than $|k|$, as it follows from Eqs. (\ref{TFA}) and (\ref{GSL}):
\begin{equation}
\left( U_{\min }\right) _{\mathrm{TF}}=(1/2)\sqrt{|k|-A}.  \label{Umin_TF}
\end{equation}
Beyond the framework of the TF approximation, $U_{\mathrm{min}}$ depends on the width of the potential barrier, $W$, naturally decreasing with the increase of $W$, as seen in Fig. \ref{fig3}(b).

Next, the dependence $U_{\mathrm{min}}(W)$ is plotted in Fig. \ref{fig3}(b), where $U_{\mathrm{min}}$ decreases with the increase of $W$, and Fig. \ref {fig3}(c) demonstrates that $U_{\mathrm{min}}$ increases (i.e., the bubble becomes more shallow) with the increase of $|k|$. In all cases, the values of $U_{\min }$ are lower for larger $A$, which is explained by the fact that the potential barrier (\ref{GSL}) with larger $A$ provides a stronger repulsive effect.

The eigenvalue spectra, examples of which are displayed in Figs. \ref{fig2}(b,e,h), demonstrate that all the bubble solutions are stable. Further, perturbed long-distance propagation (up to $z=1000$, which is tantamount to $\sim 200$ Rayleigh (diffraction) lengths of the respective stationary solutions) of the bubbles marked by S1--S3 in Figs. \ref{fig2}(c,f,i) is displayed in Figs. \ref{fig4}(a--c), respectively, which confirms the predicted stability. The Rayleigh length can be identified in Figs. \ref{fig4}(d--f), which displays the diffraction of the same inputs in the case when the potential barrier (\ref{GSL}) is removed in Eq. (\ref{NLFSE}).

\subsection{W-shaped solitons}

\label{sec3b}

Typical profiles of W-shaped dark modes, supported by the double potential barrier (\ref{WSL}), and the dependence of the respective power defect $\Delta P$ on parameters are displayed in Fig. \ref{fig5}. It is observed that the dependences are qualitatively similar to their counterparts for the bubbles, shown above in Fig. \ref{fig2}. The profiles of the W-shaped modes
marked by S4 and S5 in Fig. \ref{fig5}(a4) are presented in Figs. \ref{fig5}(a1) and (a2), respectively, and the spectrum of the perturbation eigenvalues for the former solution is shown in panel (a3). Similarly, the profiles of the modes marked by S6 and S7 in Fig. \ref{fig5}(b4) are presented in Figs. \ref{fig5}(b1) and (b2), respectively, and the eigenvalue spectrum for the former one is shown in Fig. \ref{fig5}(b3). Next, the profiles of the W-shaped modes marked by S8 and S9 in Fig. \ref{fig5}(c4) are presented in Figs. \ref{fig5}(c1) and (c2), respectively, and the eigenvalue spectrum of the former one is presented in Fig. \ref{fig5} (c3).

In Figs. \ref{fig6}(a), (b), and (c) we plot, severally, the minimum value $U_{\mathrm{\min }}$ of the W-shaped mode as a function of $A$ for different values of $W$, as a function of $W$ for different values of $A$, and as a function of $k$ for different values of $A$. These dependence are qualitatively similar to their counterparts for the bubbles displayed in Fig. \ref{fig3}, and can be qualitatively explained by means of the same arguments that are mentioned above for the bubbles.

Similar to the bubbles, the W-shaped modes are also completely stable. This conclusion is illustrated by examples of spectra of perturbation eigenvalues presented in Figs. \ref{fig5}(a3,b3,c3). The predicted stability is confirmed by numerical simulations of the propagations of perturbed modes marked by S4, S6, and S8 in Figs. \ref{fig7}(a--c), respectively. For the sake of comparison, the diffraction of the same inputs in the absence of the double potential layer is displayed in Figs. \ref{fig7}(d,e,f).

\section{Conclusion}

\label{sec4} We have constructed the bubble and W-shaped modes, supported, respectively, by the single and double Gaussian potential barrier, in the medium with the defocusing cubic term and fractional diffraction. This was done systematically in the numerical form, and also analytically, with the help of the TF (Thomas-Fermi) and variational approximations. The modes are
characterized by the power defect $\Delta P$ and the minimum value of the field, $U_{\min }$. Their dependences on parameters of the system and propagation constant of the mode families, $k$, have been found. The computation of eigenvalues for small perturbations, as well as direct numerical simulations of the perturbed evolution, demonstrate complete stability of both the bubbles and W-shaped modes.

\section*{Funding}

National Major Instruments and Equipment Development Project of National Natural Science Foundation of China (No. 61827815); National Natural Science Foundation of China (No. 62075138); Science and Technology Project of Shenzhen (JCYJ20190808121817100, JCYJ20190808164007485, JSGG20191231144201722); Israel Science Foundation (No. 1286/17).

\section*{Conflict of interest}

The authors declare that they have no conflict of interest.

\end{document}